\documentclass[letterpaper,twocolumn,english,aps,prl,groupedaddress,superscriptaddress,showpacs,amsfonts]{revtex4}
\usepackage{ae}
\usepackage{aecompl}
\usepackage[T1]{fontenc}
\usepackage[latin1]{inputenc}
\usepackage{float}
\usepackage{amsmath}
\usepackage{graphicx}
\usepackage{amssymb}

\makeatletter
\usepackage{babel}
\makeatother
\begin{document}

\title{First-principles Calculation of the Single Impurity Surface Kondo Resonance}

\author{Chiung-Yuan Lin}

\affiliation{Department of Physics, Boston University, 590 Commonwealth Avenue,
Boston, Massachusetts 02215, USA}

\affiliation{IBM Almaden Research Center, San Jose, CA 95120-6099,
USA}

 \author{A.~H. Castro Neto}

 \affiliation{Department of Physics, Boston University, 590 Commonwealth Avenue,
Boston, Massachusetts 02215, USA}

 \author{B.~A.~Jones}

\affiliation{IBM Almaden Research Center, San Jose, CA 95120-6099,
USA}

\date{\today}

\begin{abstract}
We have performed first-principles calculation of the surface and
bulk wavefunctions of the Cu(111) surface and their hybridization
energies to a Co adatom, including the potential scattering from
the Co. By analyzing the calculated hybridization energies, we
found the bulk states dominate the contribution to the Kondo
temperature, in agreement with recent experiments. Furthermore, we
also calculate the tunneling conductance of a scanning tunneling
microscope (STM) and compare our results with recent experiments
of Co impurities in the Cu(111) surface. Good quantitative
agreement is found at short parallel impurity-tip distances (
$<6\,$\AA). Our results indicate the need for a new formulation of the
problem at larger distances.
\end{abstract}

\pacs{68.37.Ef, 72.15.Qm, 72.10.Fk}

\maketitle

Scanning tunneling microscopy (STM) has become one of the most
basic tools for the manipulation of matter at the atomic scale.
Although this experimental technique has reached maturity, the
detailed theoretical understanding of experimental data is still
incomplete and/or contradictory. One of the most famous examples
of atomic manipulation is associated with the surface Kondo effect
observed when transition metal ions (such as Co) are placed on a
metallic surface (such as Cu(111)) \cite{science}. The surface
Kondo effect is the basis for the observation of surprising
phenomena such as quantum mirages \cite{mirage}, and has attracted
a lot of attention and interest in the last few years. The current
understanding of these observations is based on the assumption
that only surface states of Cu(111) are involved in the scattering
of electron waves by the Co adatoms
\cite{Heller,mgprl,mirage-prb}. Nevertheless, recent experiments
with Co atoms on the Cu(100) surface (that does not have any
surface state) \cite{Knorr}, or in Cu(111) but close to atomic
surface steps (that affect the surface states) \cite{Limot-Berndt}
have indicated that bulk (not surface) states are behind the
surface Kondo effect. Meanwhile, in contrast, the growing
theoretical literature in the subject is heavily concentrated on
the surface states alone. In this paper, we use first-principles
methods that clearly show that the bulk states are behind the
surface Kondo effect, in agreement with these experiments. In the
light of these results, theoretical approaches based on surface
states alone have to be reconsidered.

The surface Kondo effect without the STM probe is described by the
Anderson impurity model:
\begin{eqnarray}
H_{s}&=&\sum_{{\bf k}\sigma}\epsilon_{{\bf k}}c_{{\bf
k}\sigma}^{\dagger}c_{{\bf k}\sigma}+\sum_{{\bf k}\sigma}(t_{{\bf
k}a}c_{{\bf k}\sigma}^{\dagger}c_{a\sigma}+{\rm{H}}.{\rm{c}}.) \nonumber \\
&&+\sum_{\sigma}\epsilon_{a}c_{a\sigma}^{\dagger}c_{a\sigma}
+Un_{a\uparrow}n_{a\downarrow} \, ,
\label{Anderson}
\end{eqnarray}
where $\epsilon_{{\bf k}}$ is the energy of the substrate
electrons, $c_{{\bf k}\sigma}^{\dagger}$ ($c_{{\bf k}\sigma}$)
creation (annihilation) for substrate electrons with momentum {\bf
k} and spin $\sigma$, $\epsilon_{a}$ is the energy of the adatom
$d$-orbital, $c_{a\sigma}^{\dagger}$ ($c_{a\sigma}$) is the
creation (annihilation) of adatom electrons, and $U$ is the
Coulomb energy for double occupancy of the $d$-orbital
($n_{a\sigma}$ is the number operator). The hybridization energy
between the substrate and adatom is written in terms of matrix
elements between their wavefunctions.
\begin{equation}
t_{{\bf{k}}a}=\int{d}^{3}r\,\psi_{\bf{k}}^{\ast}({\bf{r}})
(H_{0}+V_{a}({\bf{r}}))\psi_{a}({\bf{r}}), \label{Vka}
\end{equation}
where $H_{0}$ refers to the bare metal surface, $V_{a}(r)$ is the
adatom scattering potential, $\psi_{\bf{k}}({\bf{r}})$ is the
substrate wavefunction scattered by the adatom potential, and
$\psi_{a}({\bf{r}})$ the adatom $3d$ orbital. Most work on the
surface Kondo effect \cite{Ujsaghy,Plihal-Gadzuk} follows
Anderson's original idea \cite{Anderson} in dealing with the
hybridization matrix element $t_{{\bf{k}}a}$, namely, treating
it as a phenomenological parameter to fit experiments.
However, in trying to understand the STM experiments, and
especially the role played by the surface and bulk states, these
matrix elements cannot be simply taken as phenomenological
parameters since one would not be able to separate the
contributions coming from the bulk and the surface states of the
substrate. Thus, it is essential to perform microscopic
calculations of these matrix elements starting from the electronic
wavefunctions.

Microscopic calculations of these matrix elements using the nearly
free-electron model (NFE) for Cu have been attempted recently
\cite{prb,Merino-Gunnarsson}. The NFE approximation has its
advantages in analyzing the momentum dependence of the
hybridization energies and obtaining analytical substrate
wavefunctions. However, in using the NFE one needs to prescribe
how the substrate crystal potential joins its vacuum image
counterpart, and the uniqueness of this potential prescription is
questionable. Such a freedom in modeling potential makes the NFE
method unreliable in obtaining the precise surface and bulk
wavefunctions. Moreover, we further notice that the substrate
states in (\ref{Anderson}) are those already scattered by the
adatom potential rather than the bare-substrate states. The
relation between these adatom-scattered states and the bare
substrate states is given by:
\begin{eqnarray}
\sum_{{\bf k}\sigma}\epsilon_{{\bf k}}c_{{\bf
k}\sigma}^{\dagger}c_{{\bf k}\sigma} &=& \sum_{{\bf
k}\sigma}\epsilon_{{\bf k}}c_{{\bf k}\sigma}^{(0)\dagger}c_{{\bf
k}\sigma}^{(0)}
\nonumber \\
&+& \sum_{{\bf k}{\bf k}'\sigma}\left(U_{{\bf k}{\bf
k}'}c_{{\bf k}\sigma}^{(0)\dagger}c_{{\bf
k}'\sigma}^{(0)}+H.c.\right), \label{potential-scattering}
\\
U_{{\bf k}{\bf
k}'} &=& \int{d}^{3}r\,\psi_{\bf{k}}^{(0)\ast}({\bf{r}})
V_{a}({\bf{r}})\psi_{\bf{k}'}^{(0)}({\bf{r}}), \label{Ukk}
\end{eqnarray}
where the superscript $(0)$ refers to the bare electronic states.
The NFE does not consider the scattering potential from the adatom
and uses the bare-substrate states in the Anderson impurity model
(\ref{Anderson}).

In the presence of the tip new terms have to be added to the
Hamiltonian that describe the tunneling processes of tip-to-adatom
and tip-to-substrate \cite{Plihal-Gadzuk}. The tip-to-adatom
tunneling process is described by:
$H_{at}=\sum_{\sigma}t_{ap}\left(c_{a\sigma}^{\dagger}c_{p\sigma}
+{\rm{H}}.{\rm{c}}.\right)$, where $c_{p\sigma}$
($c^{\dagger}_{p\sigma}$) annihilates (creates) electrons at the
tip and $t_{ap}$ is the hybridization energy between tip and
adatom. The tip-to-substrate hybridization is given by:
$H_{st}=\sum_{k\sigma}t_{{\bf k}p}\left(c_{{\bf
k}\sigma}^{\dagger}c_{p\sigma} +{\rm{H}}.{\rm{c}}.\right)$, where
$t_{{\bf k}p}$ is the hybridization energy between substrate (bulk
or surface state) and adatom. The tip Hamiltonian is simply:
$H_{t}=\sum_{\sigma}\epsilon_p c_{p\sigma}^{\dagger}c_{p\sigma}$,
where $\epsilon_p$ is the energy of the tip electrons. The total
Hamiltonian of the tip-substrate-adsorbate system is:
$H=H_{s}+H_{t}+H_{at}+H_{st}$. For the hybridization energies that
involve the STM tip, we follow Plihal and Gadzuk
\cite{Plihal-Gadzuk} and approximate
$t_{p\alpha}\propto\psi_{\alpha}^{\ast}({\bf R}_{t})$ with
$\alpha={\bf k}$, $a$.

\begin{figure}
\begin{center}
\includegraphics[keepaspectratio,width=6.5cm]{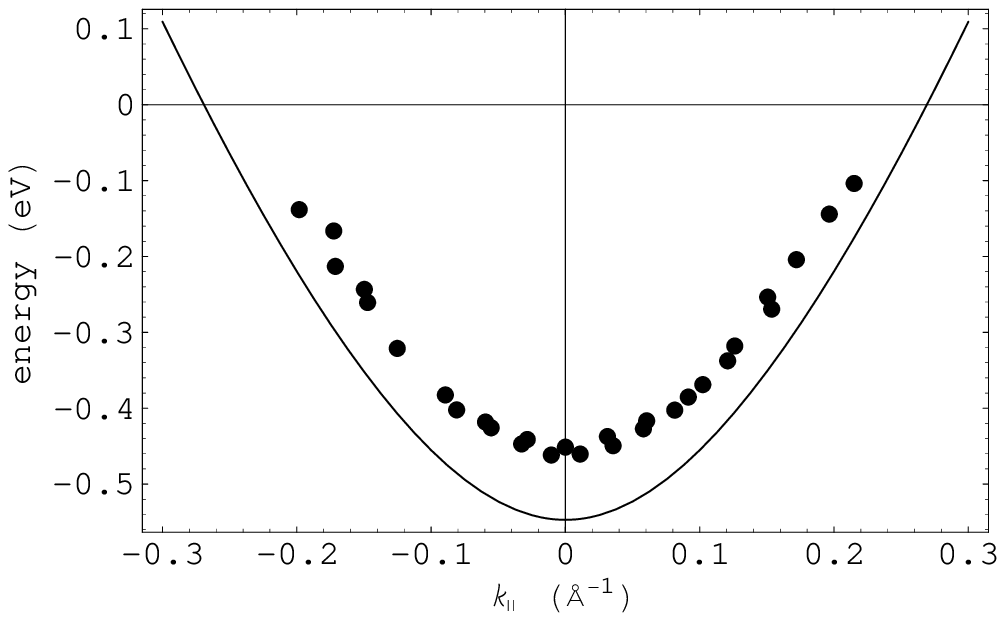}
\includegraphics[keepaspectratio,width=4.5cm]{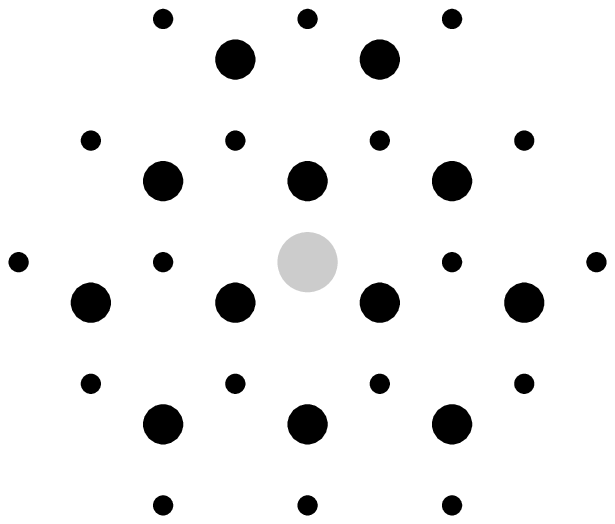}
\end{center}
\caption{\label{geometry} The left figure shows the surface-state
dispersion from our calculation (curve) and the experiments \cite{Kevan}
(data points). The right figure shows the top view of
the unit cell of Co on the Cu(111) surface in our LSDA calculation
of the present work. The Co atom is located at the center (grey
circle). The larger and smaller circles are the first (surface)
and second Cu layers, respectively. The individual Co atoms are
separated by $8\,$\AA.}
\end{figure}

To correctly obtain the contributions of the surface Kondo
resonance from the surface and bulk states, we perform
first-principles calculations of surface and bulk wavefunctions on
the Cu(111) surface in the presence of the scattering potential,
as well as their hybridization energies to the Co adatom. As the
first step, we calculate the wavefunctions of a bare Cu(111)
surface. Such a surface is simulated by a super-cell of 21-layer
slabs separated by 8 vacuum layers.
We employ, in the framework of density functional theory, a
self-consistent full-potential linearized augmented plane wave
(FLAPW) method \cite{win2k}, with the exchange-correlation
potential in the generalized gradient approximation (GGA)
\cite{PBE96}. The interatomic distances within a Cu slab are
determined by the bulk lattice constant $a_{0}=3.62\,$ \AA, and the
surface relaxations of Cu(111) are neglected because of its
close-packed
structure.
The calculated surface-state dispersion
agrees very well with the experiments (Fig.~\ref{geometry} left).

In order to obtain the potential of a Co adatom, we perform another
FLAPW calculation, in the local spin density approximation (LSDA),
of 7-layer Cu slabs separated by 8 vacuum layers, plus Co
impurities 1-layer spacing high on top of the Cu surface layers
(Fig.~\ref{geometry} right). The potential
$V_{a}({\bf{r}})$ appearing in both (\ref{Vka}) and (\ref{Ukk}) is taken to be
the potential difference between the surface with Co and the clean
crystal. Using the bare Cu(111) states within the energy range
$|\epsilon-\epsilon_{F}|<1$ eV as basis, we perform an exact matrix
diagonalization on the right-hand side of
(\ref{potential-scattering}) and obtain the adatom scattered
states. It is the $3d$ orbital of the Co adatom that actually
participates in the Kondo resonance. The $3d$ orbital
$c_{a\sigma}$ appearing in (\ref{Anderson}) is renormalized by the
Cu-substrate potential but does not include hybridization with the
substrate states. Our current first-principles approach cannot
generate $3d$ orbitals satisfying both conditions. Instead we can
perform LSDA calculations for the $3d$ orbital of either a single
Co atom or a Co atom on Cu(111) with hybridization $t_{a {\bf k}}$
included. The latter is not a good candidate to be used in
(\ref{Anderson}) because it adsorbs $t_{a {\bf k}}$ into itself
and should in principle give $t_{a {\bf k}}\approx0$. Thus we
calculate the electronic structure of a single Co atom using a
relativistic atomic code \cite{Desclaux} and use its
$3d_{3z^{2}-r^{2}}$ orbital as the $\psi_{a}({\bf{r}})$ in
(\ref{Vka}). The particular choice of the $3d$ orbital
$3d_{3z^{2}-r^{2}}$ other than $3d_{xy}$ or $3d_{x^{2}-y^{2}}$ is
supported by Ref.~\cite{Nordlander-Tully}.

The theory of the surface Kondo resonance adopted in the present
work closely follows Ref. \cite{prb,Newns&Read}. The broadening of
the Co $d$-level is calculated directly from $t_{{\bf{k}}a}$ in (\ref{Vka}) 
(without any adjustable parameters)
\begin{equation}
\Delta=\pi\sum_{\bf k} |t_{{\bf
k}a}|^{2}\delta(\epsilon_F-\epsilon_{{\bf k}}), \label{Delta}
\end{equation}
where $\epsilon_F$ is the Fermi energy, to be $\Delta=0.18$ eV. In
fact, the STM-measured Kondo temperature for Co/Cu(111) is
$T_{K}\sim 50K$, which, from
$T_{K}=D\exp\left(-\pi|\epsilon_{a}|/\Delta\right)$, gives
$\Delta\sim0.2$ eV if one uses well established values of the $d$
level $\epsilon_{a}\sim 0.9$ eV \cite{Ujsaghy} and the Cu band
cutoff $D\sim5.5$ eV \cite{Smith}. The contributions to $\Delta$
from the surface and bulk states are also investigated by
considering $\Delta=\Delta_{\rm surf}+\Delta_{\rm bulk}$. We
calculate from (\ref{Delta}) the ratio of the surface-state
contribution to the total $d$-level broadening $\Delta_{\rm
surf}/\Delta\sim0.006$. The ratio shows that the bulk states
dominate the adatom-substrate hybridization energy of Co/Cu(111).
This result can be understood by the fact that the Co atom is
still in one layer above the surface Cu layer, a crystal-like
regime rather than the tunneling regime (that is, the Co adatom is
strongly hybridized with the substrate).

The STM differential conductance can be written as:
\begin{equation}
\frac{dI}{dV}-\left(\frac{dI}{dV}\right)_{0}= a(R)
\frac{\left|q(R)\right|^{2}-1+2\, \xi \, {\rm Re}\,[q(R)]}{\xi^{2}+1}.
\label{dIdV-generalized}
\end{equation}
where $R$ is the parallel impurity-tip distance,
$\xi=(eV+\bar{\epsilon}_{a})/(k_{B}T_{K})$ is the
dimensionless bias, $\bar{\epsilon}_{a}$ is a bias off-set
due to the $d$-state energy,
\begin{eqnarray}
a(R)= \left|
\pi \sum_{{\bf k}}
 t_{p{\bf k}} t_{{\bf k}a} \delta(\epsilon_F - \epsilon_{{\bf k}})
 \label{ar}
 \right|^2 \, ,
\end{eqnarray}
is the amplitude of the resonance, and
\begin{equation}
q=\frac{t_{pa}+ \sum_{{\bf k}} t_{p{\bf k}}t_{{\bf k}a}{\rm{Re}}\,G_{\bf k}}
{\sum_{{\bf k}} t_{p{\bf k}}t_{{\bf k}a}{\rm{Im}}\,G_{\bf k}}
\label{qfull}
\end{equation}
is the so-called Fano parameter where $G_{\bf
k}=(\epsilon_{F}-\epsilon_{{\bf k}}+i\eta)^{-1}$ is the bare
substrate Green's function. In (\ref{dIdV-generalized}), the
differential conductance with a subscript ``0" refers to the
background signal (proportional to the local density of states of
the substrate). Typically $q(R)$ has been taken to be real
in the experimental fits.
However, when performing the first-principles calculation of the Cu(111)
surface, we found that $q$ can carry an imaginary part.
The Bloch states of the conduction electrons can 
generally be written as $\psi_{n{\bf k}}({\bf r})=u_{n{\bf
k}}({\bf r})e^{-i{\bf k}_{\parallel}\cdot{\bf r}_{\parallel}}$
with $u_{n{\bf k}}({\bf r})=\left|u_{n{\bf k}}({\bf
r})\right|e^{i\Phi({\bf r})}$. NFE studies of the surface Kondo
resonance \cite{Plihal-Gadzuk,prb,Merino-Gunnarsson} treat
$u_{n{\bf k}}$ in approximation such that $\Phi({\bf r})$ is spatially
independent, i.e., $\Phi({\bf r})=\Phi_{0}$ is an overall phase.
However, our first-principles calculation shows that the function
$u_{n{\bf k}}$ can carry a spatially varying phase. Including
$\Phi({\bf r})$ in the calculation of the hybridization energies
consequently gives the complex
$q$.

The STM tip is positioned about $5\,$\AA $<z<$ $10\,$\AA \ above the
surface in the usual spectroscopic tunneling conditions. However,
when performing first-principles calculations of the bare-Cu(111)
wavefunctions, we find that the wavefunctions undergo an
oscillatory behavior rather than a smooth exponential decay beyond
$4\,$\AA \ from the surface. It is known that this problem comes from
GGA in the low density region because of large-scaled gradients \cite{GGAwfn}.
The large gradients cause fluctuations in the exchange-correlation potential, which
leads to
fluctuations in
wavefunctions. To resolve this problem we extrapolate the
Cu-substrate wavefunctions by fitting their values in $2\,$\AA$<z<3\,$\AA
\ using the NFE wavefunctions \cite{prb}. The NFE wavefunction
is then used in the $z>5\,$\AA \ region to calculate the STM
differential conductance.

\begin{figure}
\begin{center}
\includegraphics[keepaspectratio,width=6.5cm]{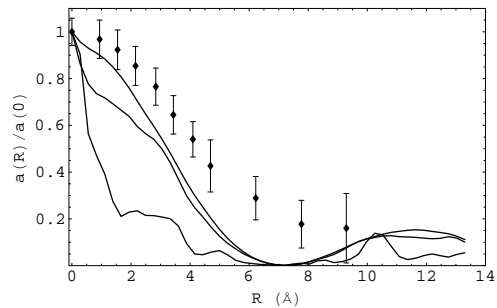}
\end{center}
\caption{\label{fig-amplitude-3heights} Normalized Fano formula
prefactors $a(R)/a(0)$ at three tip heights $Z_{t}=3.5\,$\AA,
$10\,$\AA, $16\,$\AA \ (from the bottom to the top). The experimental
data is also shown \cite{Knorr}.}
\end{figure}

We analyze surface and bulk-state contributions to $a(R)$, and
find that the bulk states dominate. In
Fig.~\ref{fig-amplitude-3heights} we plot $a(R)$ of three tip
heights ($Z_{t}=3.5\,$\AA, $10\,$\AA, $16\,$\AA) as well as the
experimental data \cite{Knorr}. The prefactor $a(R)$ at
$Z_{t}=3.5\,$\AA \ is calculated directly from the GGA
wavefunctions rather than the extrapolated NFE wavefunctions. It
is clear that $a(R)$ moves towards the
experimental data as the tip moves farther away from the surface.
One can see that our first-principles calculation at $Z_{t}=16\,$\AA
\ (without fitting parameters) agrees well with the experiment
within $R<5\,$\AA \ but starts deviating from the experiments as
$R$ increases and has a node around $R=7\,$\AA. Since the current
first-principles approach in calculating the Cu(111) electronic
structure, GGA, is widely regarded as a highly accurate
computation scheme except for very low electron density,
we suggest that further theoretical and
experimental work is required to check this issue. In the bulk
Kondo problem the conduction-impurity hybridization can be
regarded as momentum-independent, and the Kondo Hamiltonian has an
exact solution by Bethe ansatz \cite{Bethe-ansatz}. However, the
surface Kondo effect has a ${\bf k}$-dependent $t_{{\bf k}a}$, and
there is so far no field-theoretical approach that can treat it
exactly. A possible solution is to use a computational scheme to
compute the Anderson Hamiltonian of a surface Kondo effect. In the
experimental aspect it should be pointed out that at large values
of $R$ the STM data is very noisy, and it is possible that the
fitting is not unique.

\begin{figure}
\begin{center}
\includegraphics[keepaspectratio,width=6.5cm]{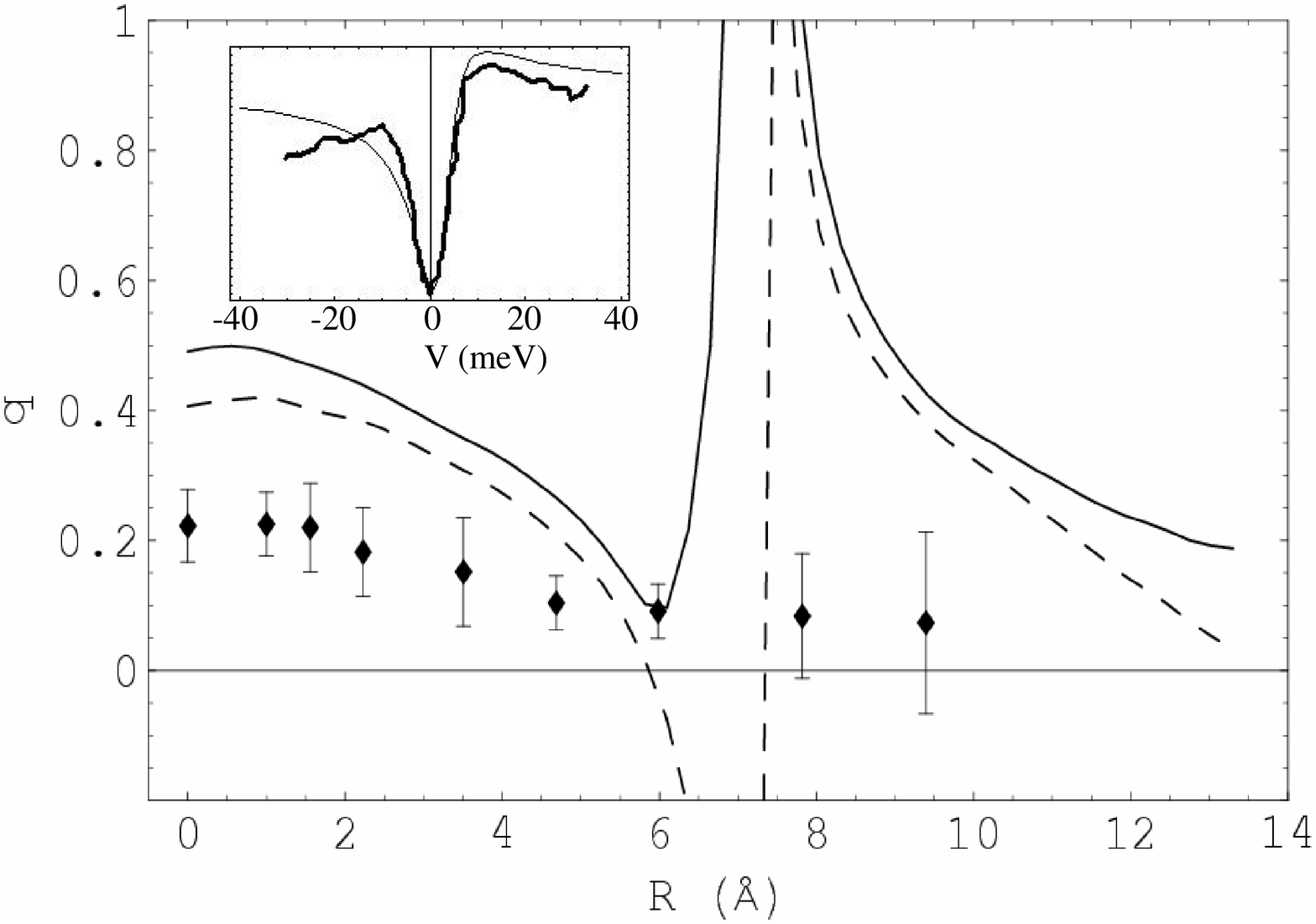}
\end{center}
\caption{\label{fig-q}The absolute value (solid curve) and real
part (dashed curve) of  calculated complex $q$ at $Z_{t}=16\,$\AA \
as a function of $R$ as compared with the experimental fit
assuming ${\rm Im}[q]=0$. Inset: comparison of theoretical (light) and experimental
(dark) STM $dI/dV$ vs.~$V$ lineshapes at $R=0$.}
\end{figure}

The calculation of $a(R)$ requires only the Cu-substrate states at
the Fermi energy (see Eq.~(\ref{ar})) while the lineshape
parameter $q$ in (\ref{qfull}) depends on the entire Cu band. We
use the calculated Co-scattered Cu states within the energy range
$|\epsilon-\epsilon_{F}|<0.9$ eV to obtain the $q(R)$ in
Eq.~(\ref{qfull}). Since the Fano lineshape parameter $q$ defined
in (\ref{qfull}) is in general complex, the experimentally fitted
$q$ based on the Fano formula of (\ref{qfull}) with ${\rm
Im}[q]=0$ can only be compared with our calculation qualitatively.
We plot our calculated $|q|$, Re~$q$, and the experimentally
fitted $q$ in Fig.~\ref{fig-q}. The inset is a direct comparison of
between our calculated and the experimental STM lineshapes at $R=0$,
showing good agreement.
One can see likewise that the $q$ vs.~$R$ plot shows good agreement
between our calculation and the experiments for $R<6\,$\AA \ and
the discrepancy for $R>6\,$\AA \ is a consequence of the node of
$a(R)$ around $R=7\,$\AA.

The effect of the potential scattering from the Co atom is also
studied by calculating the $d$-level broadening $\Delta$ and the
STM lineshape prefactor $a(R)$ without potential scattering,
i.e., $U_{{\bf k}{\bf k}'}=0$ in (\ref{potential-scattering}).
When potential scattering is neglected, the $d$-level
broadening $\Delta$ reduces by $8\%$, and its surface-state
contribution slightly increases but is still small
($\Delta_{\rm{surf}}/\Delta\sim0.025$). We also found that without
potential scattering the surface states dominate the
contribution to the STM lineshape prefactor $a(R)$. This is a
drastic change from the potential scattering case where bulk
states dominate. This change can be understood as
follows: bulk states dominate the local density of states (LDOS)
of the conduction electrons at the adatom site with and without potential
scattering, which accounts for the $d$-level broadening. As LDOS
of the bulk states decays away from the surface faster than the
surface states, in the case of no potential scattering the product
of $t_{ak}$ (bulk dominated) and $t_{pk}$ (surface dominated) in
(\ref{ar}) turns out to be dominated by the surface states. When
potential scattering is included, the bare-substrate bulk and
surface states strongly mix with each other. As a result, LDOS of the bulk
and surface states in the presence of potential scattering
decay from the surface approximately in the same rate, and
$t_{ak}t_{pk}$ becomes dominated by the bulk states.

In summary, we have calculated the hybridization energies from the
LDA wavefunctions of the Cu (111) surface and Co atom. The
potential scattering of Cu conduction states from the Co adatom is
included in determining the substrate-adatom hybridization energy.
Our calculated $d$-level broadening from the above hybridization
energy is in excellent agreement with the value determined from
the STM-measured Kondo temperature. Our analysis of the
contribution of the substrate-adatom hybridization energy from
surface and bulk states shows that the bulk states dominate the
Kondo temperature. We also calculated the tunneling conductance of
an STM tip for the Cu(111) surface in the presence of a Co adatom.
Our calculated conductance has quantitative agreement with the
experiments at short parallel tip-adatom distance without any
adjustable parameters. However, discrepancy appears as the
parallel distance increases indicating that a new approach is
required for this problem.

{\bf Acknowledgments} We would like to thank O.~Gunnarsson,
A.~Heinrich, L.~Limot, V.~Madhavan, J.~Merino, N.~Sandler, and
A.~Zawadowski for illuminating discussions. We would like to
acknowledge support under DARPA contract no.~DAAD19-01-C-0060.
A.H.C.N. was partially supported through NSF grant DMR-0343790.

\end{document}